\documentclass[aps,prx,twocolumn,longbibliography]{revtex4-2}
\usepackage{graphicx}
\usepackage{dcolumn}
\usepackage{bm}
\usepackage{amsmath}
\usepackage{amsfonts}
\usepackage{graphicx}
\usepackage{color}

\def\bra#1{\mathinner{\langle{#1}|}} 
\def\ket#1{\mathinner{|{#1}\rangle}}

\newcommand{\Eq}[1]{Eq. (\ref{#1})}

\begin{document}

\title{Theoretical proposals to measure resonator-induced modifications of the electronic ground-state in doped quantum wells}

\author{Yuan \surname{Wang}}
\author{Simone \surname{De Liberato}}
\email[Corresponding author: ]{s.de-liberato@soton.ac.uk}
\affiliation{School of Physics and Astronomy, University of Southampton, Southampton, SO17 1BJ, United Kingdom}

\begin{abstract}
Recent interest in the physics of non-perturbative light-matter coupling led to the development of solid-state cavity quantum electrodynamics setups in which the interaction energies are comparable with the bare ones. In such a regime the ground state of the coupled system becomes interaction-dependent and is predicted to contain a population of virtual excitations which, notwithstanding having been object of many investigations, remain still unobserved. In this paper we investigate how virtual electronic excitations in quantum wells modify the ground-state charge distribution, and propose two methods to measure such a cavity-induced perturbation. 
The first approach we consider is based on spectroscopic mapping of the electronic population at a specific location in the quantum well using localised defect states. The second approach exploits instead the photonic equivalent of a Kelvin probe to measure the average change distribution across the quantum well.
We find both effects observable with present-day or near-future technology. Our results thus provide a route toward a demonstration of cavity-induced modulation of ground-state electronic properties. 
\end{abstract}

\maketitle

\section{Introduction}
\label{Introduction}
The many advances in the fabrication of nanophotonic resonators and nanostructured materials have made solid-state cavity quantum electrodynamics (CQED) an interdisciplinary research domain, with applications ranging from chemistry \cite{feist_polaritonic_2018} to machine learning \cite{ballarini_polaritonics_2019}. One of the figures of merit of CQED setups which has seen sustained improvements is the coupling strength between light and matter. When their mutual interaction energy becomes comparable with the excitation energy, higher order effects become observable, a regime called ultrastrong coupling \cite{de_bernardis_cavity_2018,frisk_kockum_ultrastrong_2019,forn-diaz_ultrastrong_2019}. In 2009  the impact of these higher-order perturbative effects was observed for the first time in the anomalous shift of intersubband polariton resonances: transitions between conduction subbands in doped semiconductor quantum wells (QWs) strongly coupled to photonic resonators \cite{anappara_signatures_2009}. 
Large couplings with the photonic vacuum is expected to modify not only the system's optical response, but also the underlying matter degrees of freedom.
Interest in the impact of CQED effects on electronic and molecular degrees of freedom 
dates back to Khurgin's idea of very strong coupling \cite{khurgin_excitonic_2001,yang_verification_2015,brodbeck_experimental_2017,khurgin_pliable_2018}, but only in more recent years a broader interest followed \cite{galego_cavity-induced_2015,cwik_excitonic_2016,cortese_collective_2017,schafer_ab_2018,levinsen_microscopic_2019, ebbesen_hybrid_2016,herrera_cavity-controlled_2016,galego_many-molecule_2017,ruggenthaler_quantum-electrodynamical_2018,ribeiro_polariton_2018,thomas_tilting_2019,
cortese_excitons_2020}.

Crucially for this work, the ground state of an ultrastrongly coupled CQED system is predicted to host a cloud of virtual excitations, both photonic and matter ones \cite{ciuti_quantum_2005}. These excitations are predicted to be stable also in realistic dissipative environments \cite{de_liberato_virtual_2017} but, notwithstanding many theoretical works proposing different approaches to observe them, a direct measurement is still missing.
Except a proposal to observe the virtual excitations via electro-optical sampling \cite{benea-chelmus_electric_2019}, whose efficacy has been questioned \cite{de_liberato_electro-optical_2019}, and one to use the Lamb shift of an ancilla qubit 
\cite{lolli_ancillary_2015}, all the other proposals we are aware of deal with variants of one basic idea: to non-adiabatically modulate the system in order to make some of these excitations real \cite{de_liberato_quantum_2007,carusotto_back-reaction_2012,stassi_spontaneous_2013,garziano_switching_2013,huang_photon_2014,cirio_multielectron_2019,falci_ultrastrong_2019}. One problem with this idea is that we need to consider a modulated, time-dependent system, in analogy with the dynamical Casimir effect \cite{dodonov_current_2010,wilson_observation_2011,nation_colloquium_2012}, and similarly sensitive to the density of dressed states, not to the presence of vacuum excitations. 
A second problem is that, in order to achieve non-vanishing emission, perturbation frequencies of the order of the bare optical frequency are necessary \cite{gunter_sub-cycle_2009,halbhuber_non-adiabatic_2020}, a requirement which has until now thwarted any attempt to observe vacuum excitations.

This paper explores a novel approach to these problems, by noticing that 
cavity-induced virtual electronic excitations do modify the ground-state charge distribution, and they can thus be measured exploiting already well-tested methods used to map charge distributions in nanoscopic systems without requiring any non-adiabatic modulation.
Our results thus provide both a route to a first direct observation of virtual excitations in the ultrastrong-coupling vacuum, and a test bench for theoretical and numerical approaches studying CQED-induced modifications of the electronic ground state.

\section{System description}
\label{Theory}
\begin{figure}[t!]
\begin{center}
 \includegraphics[width=0.47\textwidth]{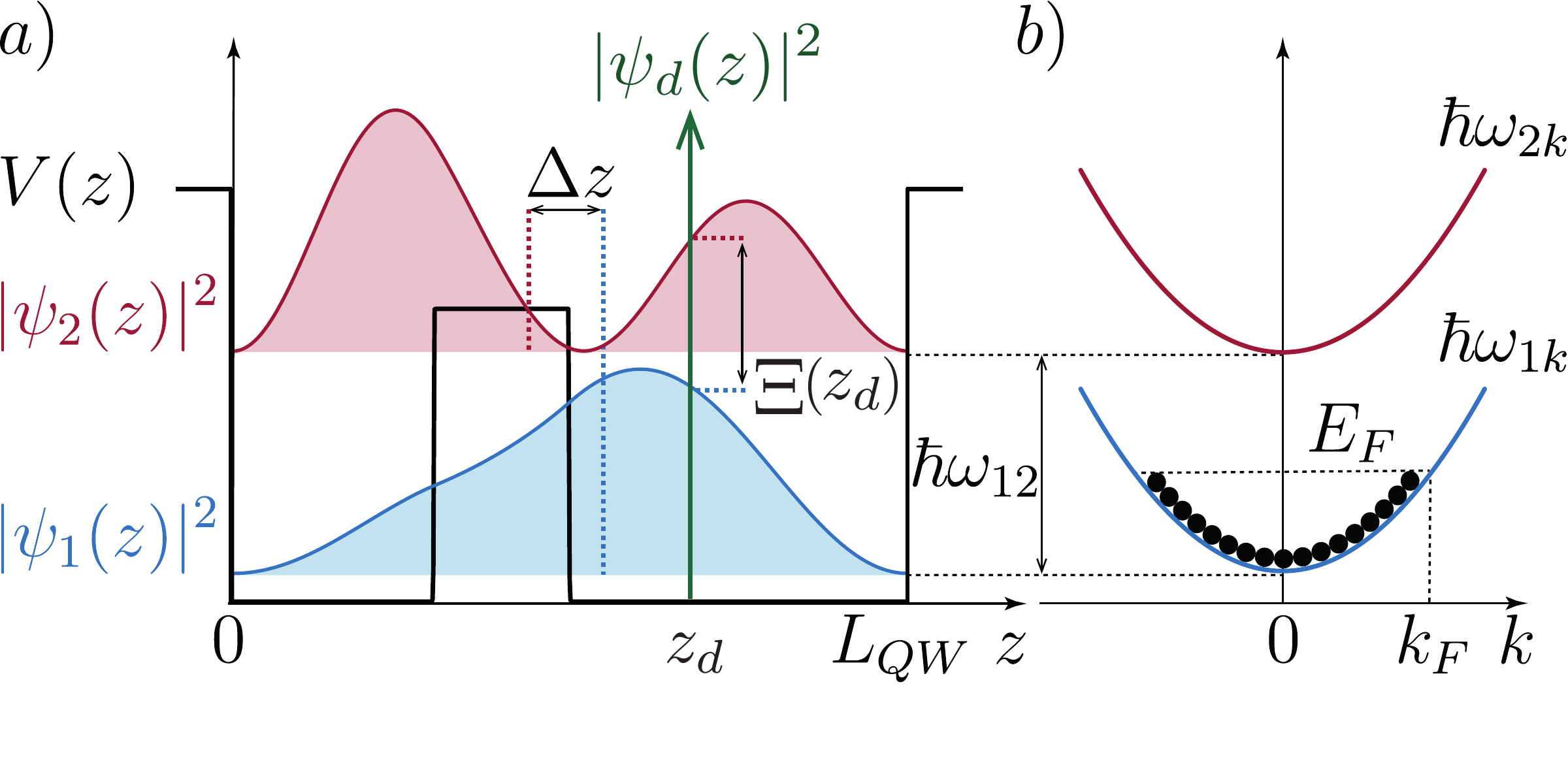}
\caption{Example of an asymmetric QW potential $V(z)$ with its envelope wavefunctions (a) and in-plane dispersions (b) of the first two conduction subbands. The vertical dotted lines represent the average charge position in each subband. 
The green arrow shows the localised defect wavefunction and the horizontal dotted lines the envelope wavefunction at the defect position in the two subbands.
The definition of all the marked quantities can be found in the main text.
\label{Fig1}}
\end{center}
\end{figure}

We consider a stack of $n_{QW}$ QWs, doped with  a two-dimensional electron gas of density $\sigma_{e}$. The electrons in each QW occupy parallel parabolic subbands with dispersion $\omega_{jk}$, function of the subband $j$ and in-plane wavevector $\mathbf{k}$. The Fermi wave vector $k_F$ is chosen such that the Fermi energy $E_F$ lies between the first and the second subbands.
Electronic wavefunctions, dispersions, and other quantities introduced in this Section are shown in Fig. \ref{Fig1}. 

The wavefunctions in th $n$th QW can be written in the  envelope function approximation
\begin{align}
\label{electron_wavefunction}
\phi_{jn\mathbf{k}}(\mathbf{r})&= \psi_j(z-z_n)\frac{e^{i\boldsymbol{k} \cdot\boldsymbol{\rho}}}{\sqrt{S}},
\end{align}
where $\mathbf{r}=(z,\boldsymbol{\rho})$ is the position vector decomposed in cylindrical coordinates, $S$ is the sample surface, and $z_n$ the $n$th QW position with the origin fixed at $z_1=0$.
We introduce the corresponding annihilation operators $\hat{c}_{jn\mathbf{k}}$, obeying Fermionic anticommutation rules 
\begin{align}
\left\{\hat{c}_{jn\mathbf{k}},\hat{c}^{\dagger}_{j'n'\mathbf{k}'} \right\}=\delta_{jj'}\delta_{nn'}\delta(\mathbf{k}-\mathbf{k}').
\end{align}
Here and in the following the electron spin is not explicitly marked and the spin multiplicity is implicitly summed over.
The ground state of the uncoupled system, describing the electron gas in the electromagnetic vacuum is then 
\begin{align}
\label{F} 
\ket{F}&=\prod_{n=1}^{n_{QW}}\prod_{k<k_F}\hat{c}^{\dagger}_{1n\mathbf{k}}\ket{0},
\end{align}
where $\ket{0}$ is the vacuum state $\hat{c}_{jn\mathbf{k}}\ket{0}=0$.
The unperturbed charge distribution corresponding to the uncoupled state $\ket{F}$ can be written as
\begin{align}
\label{rhoF}
\rho_F(z)&=\sum_n \lvert \psi_1(z-z_n)\rvert^2 \sigma_e.
\end{align}

We use the formalism initially developed for the multi-subband case \cite{todorov_dipolar_2015}, in order to be able to treat the most general cases, but for the sake of clarity in this paper we will specialise it to consider only the first two electronic subbands, an approximation which, in the proper gauge, can be justified when higher-lying subbands are either detuned or weakly coupled \cite{de_bernardis_breakdown_2018}.  

We consider the intersubband transitions to be almost vertical in momentum space.
This implies that the frequency of the resonant transition with in-plane momentum $\mathbf{q}$ is given by
\begin{align}
\label{w12}
\omega_{2\mathbf{k+q}}-\omega_{1\mathbf{k}}\approx\omega_{12}, 
\end{align}
thus neglecting terms of the order of $q v_F$, with $v_F$ the Fermi velocity. 
The QWs are embedded in a double metal nanopatch resonator of length $L$ of which we retain only transverse-magnetic (TM) modes due to selection rules of intersubband transitions. The normalised TM$_m$ mode profile can be written as 
\begin{align}
f_{m\mathbf{q}}(\mathbf{r})&=\sqrt{\frac{2}{(1+\delta_{m0})LS}}\cos(\frac{\pi m z}{L})e^{i\mathbf{k}\cdot\boldsymbol{\rho}},
\label{photon_wavefunction}
\end{align} 
with in-plane wavevector $\mathbf{q}$, out-of-plane one $\frac{\pi m}{L}$, and $m\in\mathbb{N}_0$.  The $\delta_{mn}$ is the Kronecker symbol.
Photons in these modes will be described by the boson annihilation operators $\hat{a}_{m\mathbf{q}}$, whose bare frequencies $\omega_{mq}$ obey the  dispersion relation
\begin{align}
\epsilon_r\frac{\omega_{mq}^2}{c^2}&=q^2+\frac{\pi^2 m^2}{L^2},
\label{photon_dispersion}
\end{align}
with $\epsilon_r$ the background dielectric constant and $c$ the speed of light.

We are interested to determine the modification in the charge distribution along the $z$-axis induced in the electron gas by the coupling to the photonic resonator. As such we can employ a theory whose degrees of freedom are not the electron themselves, but their transitions and in particular those coupled with the photonic field. We thus introduce the collective, bright intersubband transition in QW $n$, with in-plane wave vector $\mathbf{q}$
\begin{align}
\hat{b}_{n\mathbf{q}}^{\dagger}&=\frac{1}{\sqrt{S\sigma_{e}}}\sum_{\mathbf{k}}\hat{c}^{\dagger}_{2n\mathbf{k+q}}\hat{c}_{1n\mathbf{k}}.
\end{align}
These excitations obey quasi-bosonic commutation relations 
\begin{align}
\label{bcom}
\left[ \hat{b}_{n\mathbf{q}},\hat{b}_{n'\mathbf{q}'}^{\dagger}\right]&=\delta_{nn'}\delta(\mathbf{q}-\mathbf{q}')+\delta_{nn'}O(\frac{\sigma_n}{\sigma_{e}}),
\end{align}
and they can thus be approximated as bosons 
in the sector of the Hilbert space where the excitation density in the $n$th QW $\sigma_n$ is much smaller than the electron density 
\cite{de_liberato_stimulated_2009}
\begin{align}
\label{ssmall}
\sigma_n\ll\sigma_e.
\end{align}

The coupling of intersubband transitions to the electromagnetic field can be described using the Power-Zienau-Woolley (PZW) Hamiltonian including self-interaction terms leading to the  depolarization shift \cite{grieser_depolarization_2016}. As shown in Appendix A, such an Hamiltonian can be diagonalised in the bosonic regime using a Hopfield-Bogoliubov rotation. The coupled theory is then described by free bosonic polaritonic operators, linear superpositions of the bare ones
\begin{align}
\label{pinab}
\hat{p}_{s\mathbf{q}}&=\sum_m \left(x_{smq} \hat{a}_{m\mathbf{q}}+z_{smq} \hat{a}^{\dagger}_{m\mathbf{-q}}\right)\\&+\sum_n\left(
y_{snq}\hat{b}_{n\mathbf{q}}+w_{snq}\hat{b}_{n\mathbf{-q}}^{\dagger}\right).\nonumber
\end{align}
The linear transformation in \Eq{pinab} can then
be inverted to yield
\begin{align}
\label{binp}
\hat{b}_{n\mathbf{q}}&=\sum_s \left(\bar{y}_{snq}\hat{p}_{s\mathbf{q}}-\bar{w}_{snq}\hat{p}^{\dagger}_{s-\mathbf{q}}\right).
\end{align}
From \Eq{binp} we can calculate the total density of matter excitations in the $n$th QW in the coupled ground state defined by $\hat{p}_{s\mathbf{q}}\ket{G}=0$
\begin{align}
\label{NG}
\sigma_{n}&=\frac{1}{S}\sum_{\mathbf{q}}\bra{G}\hat{b}_{n\mathbf{q}}^{\dagger}\hat{b}_{n\mathbf{q}}\ket{G}=\frac{1}{S}\sum_{s\mathbf{q}} \lvert w_{snq}\rvert^2.
\end{align}

Intersubband transitions modify the electronic density because, as clearly shown in the example in Fig. \ref{Fig1}, the electronic charge density in each subband is different. In Appendix B we extend the approach from Ref.~\cite{cortese_strong_2019} to express the the ground state electronic wavefunction in term of bosonised excitations. We are then able to  put the ground state electronic distribution in the form
\begin{align}
\label{rz}
\rho_G(z)&=\rho_F(z)+\sum_n\Xi(z-z_n) \sigma_n,
\end{align}
with
\begin{align}
\label{Xi}
\Xi(z)&=\left[\lvert\psi_2(z)\rvert^2-\lvert\psi_1(z)\rvert^2\right].
\end{align}
In the following we will discuss two different experimental procedures to measure the quantity in \Eq{NG}, exploiting two different observables linked to the modifications it induces on the charge distribution. On the one hand, electronic excitations modify the charge distribution at a specific position $z$, as shown in \Eq{rz}. On the other hand, if the QW potential is asymmetric, electrons in different subbands have a different average positions along the sample growth axis $z$, leading to an average charge displacement when an electron jumps between the first two subbands. 
In the two-subband approximation we are employing the level of asymmetry can be quantified by a single parameter: the average electron displacement induced by an intersubband transition
 \begin{align}
 \Delta z&= \bra{\psi_2}\hat{z}\ket{\psi_2}-\bra{\psi_1}\hat{z}\ket{\psi_1}.
 \label{dz}
 \end{align}
 A graphical illustration of both $\Xi(z)$ and $\Delta z$ can be found in Fig. 1.

An estimate of the expected excitation density, to the leading order in the coupling, can be obtained by performing a lowest order expansion of $\sigma_n$ as a function of the plasma frequency of the electron gas in the QW $\omega_P$, which quantifies the strength of the collective light-matter coupling. To the first perturbative order the coupled ground state can be written as
\begin{align}
\ket{G}&\approx \ket{F}-\sum_{mn\mathbf{q}}\frac{\omega_PK_{mnq}}{\omega_{12}+\omega_{mq}}\hat{a}^{\dagger}_{m-\mathbf{q}}\hat{b}^{\dagger}_{n\mathbf{q}}\ket{F},
\end{align}
where the dimensionless coefficient $K_{mnq}$, defined in Appendix A, embeds all the other microscopic details of the CQED setup.
From \Eq{NG}, to the lowest non-trivial order, we thus obtain the following expression for the electronic excitation density in the $n$th QW
\begin{align}
\label{sGf}
{\sigma}_{n}&\approx \frac{1}{S}\sum_{m\mathbf{q}}\frac{\omega_P^2K_{mnq}^2}{(\omega_{12}+\omega_{mq})^2}.
\end{align}
The sum in \Eq{sGf} is mathematically divergent, also due to the dipolar approximation used in our theory which breaks down for photonic wavelengths of the order of the QW width.
Still, for $\omega_{mq}$ larger than a much smaller cut-off of the order of the plasma frequency of the metallic mirrors $\omega_C$, the mirrors become  transparent and their position can not efficiently affect the electromagnetic confinement. 
As we will see in the following Sections, the two measurement schemes we will propose aim to measure the impact of the cavity length on the excitation density, while filtering out the potentially much larger signal independent from the cavity length. It will thus be useful to partition the sum in \Eq{sGf} into two components
\begin{align}
\label{ss}
{\sigma}_{n}\approx {\sigma}_n^<+\sigma_n^>.
\end{align}
The first component of \Eq{ss} takes into account the sum of the modes well confined by the photonic cavity and it thus depends upon the cavity length $L$
\begin{align}
\label{<}
{\sigma}_{n}^<&= \frac{1}{S}\sum_{m\mathbf{q}}\frac{\omega_P^2K_{mnq}^2}{(\omega_{12}+\omega_{mq})^2}\Theta(\omega_C-\omega_{mq}),
\end{align}
where $\Theta$ is the Heaviside function.
The second component of \Eq{ss} includes instead all unconfined modes,
and it can be considered independent of $L$
\begin{align}
\label{sGf3}
\partial_L \sigma_n^>&=0.
\end{align}
We will develop all our theory assuming zero temperature, which is a good approximation even at room temperature for intersubband energies $\hbar\omega_{12}\gg 25$meV. Spurious thermal excitations will be present at any non-zero temperature, but their density will be independent of $L$ and we will thus operationally consider them as included in the $\sigma_n^>$ component.

For the sake of simplicity and generality in the following we will retain only the contribution of the first cavity mode $m=0$ to $\sigma_n^<$.  We can see from \Eq{sGf} that the total population of virtual excitations is a sum of positive terms corresponding to each value of $m$. Considering only the $m=0$ mode provides thus a lower bound on the intensity of the signal to be measured, allowing us to fix the minimal experimental requirements to observe it while avoiding complications due to time-variation in the number of involved photonic modes and other secondary details as the exact position of each QW in the cavity.
Inserting the coefficient $K_{mnq}$ defined in Appendix A into \Eq{sGf} we thus obtain, to the leading order in $\omega_C$ 
\begin{align}
\label{sn}
\sigma^<(L)&\approx \frac{\omega_C z_{12}^2e^2\sigma_e}{4\pi \epsilon_0 c^2 \hbar L},
\end{align}
where we marked explicitly the $L$-dependence. Given that in such an approximation the electronic excitation density is the same in each QW, in \Eq{sn} and in the following we will not mark the QW index $n$ explicitly, but it remains intended that all the electronic densities are expressed per QW.

\section{Spectroscopic measurement}
\label{Spectroscopic}
\begin{figure}[t!]
\begin{center}
 \includegraphics[width=0.5\textwidth]{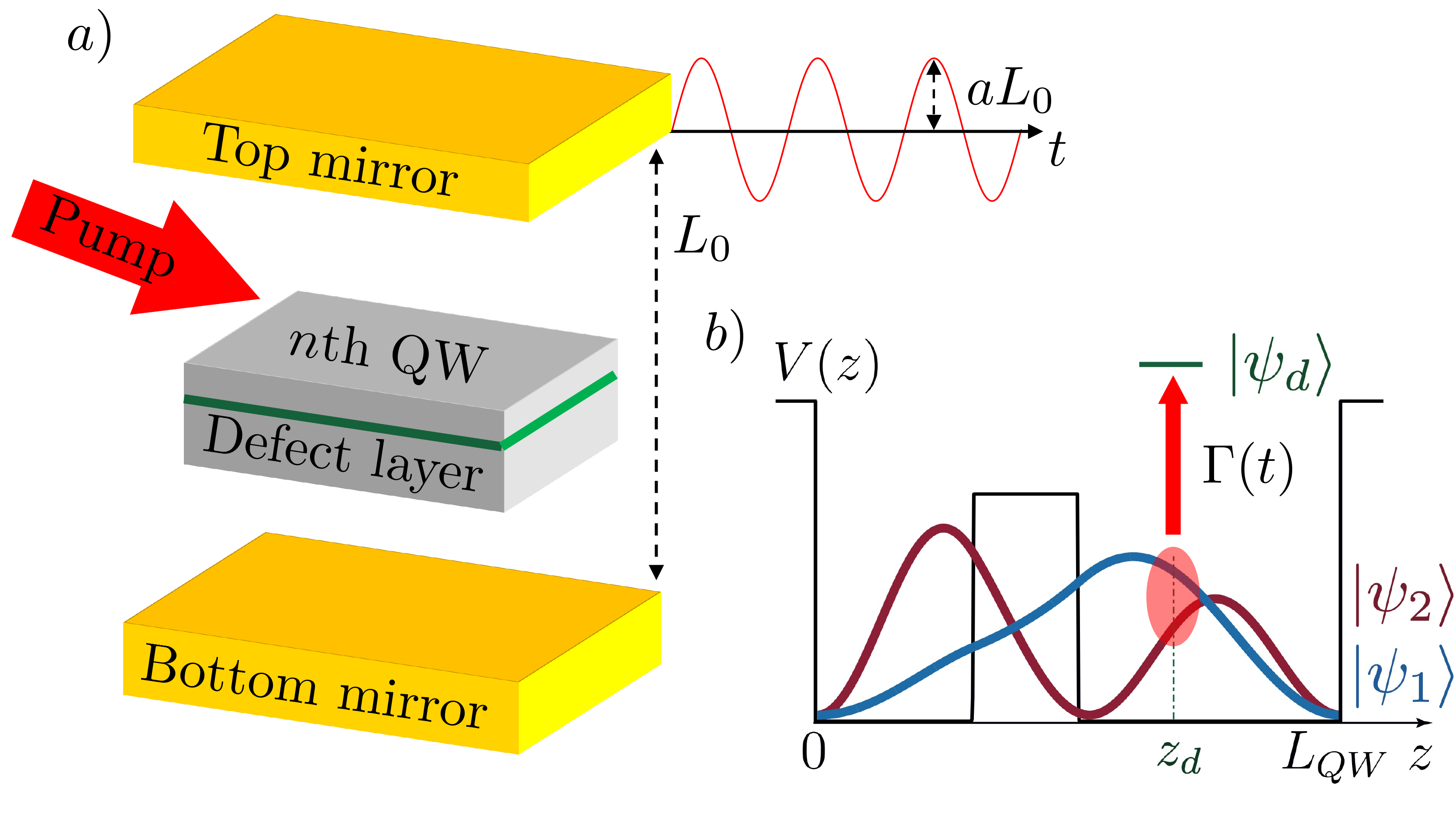}
\caption{Sketch of the setup (a) and measurement scheme (b) described in Sec. \ref{Spectroscopic}. 
The pump laser induces absorption from the two subbands into the excited defect state at position $z_d$. When the position of the top mirror is modulated in time with amplitude $aL_0$ the relative population of the two subbands, and thus the electronic density at $z_d$, is also modified, leading to a time-dependet absorption $\Gamma(t)$.}
\label{Fig2}
\end{center}
\end{figure}

In this Section we discuss how to spectroscopically measure $\sigma^<$ by using defect modes grown in each QW to map the perturbation of the electronic density. A conceptually similar procedure has already been used to measure the real-space charge density in GaAs QWs using indium and aluminum defects \cite{Marzin1989}. 
The apparatus and measurement scheme we will consider are schematised in Fig. \ref{Fig2}. The top mirror of the photonic resonator is mechanically connected to the tip of an atomic force microscope (AFM). As the tip is driven at its resonant mechanical frequency $\omega_{m}$ the cavity length $L$ is modulated around its equilibrium value $L_0$ with the normalised amplitude $a<1$
\begin{align}
\label{L}
L&=L_0\left[ 1+a\cos(\omega_{m} t)\right].
\end{align}
The oscillation changes the photonic cavity length, thus modifying the number of ground-state virtual  excitations $\sigma^<$. The mechanical resonance of the cantilever $\omega_{m}$ is normally in the kHz range, much smaller than all the other relevant frequency scales of the system. We can thus consider the driven evolution adiabatic, and the system at equilibrium in its ground state.

We consider that a planar layer of defect electronic states with energy $\omega_d$ above the first subband $\omega_{1\mathbf{k}}$ and Lorentzian lineshape with linewidth $\gamma_d$
is introduced in each QW. The defect layer is placed at the position $z_d$ corresponding to a non-vanishing value of $\Xi(z)$ in \Eq{Xi} as shown in Fig.~\ref{Fig1}(a). Note that in such a Figure we represent an explicitly asymmetric QW potential, although asymmetry does not play a role in this scheme. The asymmetry of the two subbands' wavefunctions will instead be crucial to the scheme described in the next Section, relying on a non-vanishing value of $\Delta z$.

A laser pump at frequency $\omega_d$ and power $P$ is then shone on the system, leading to electrons being excited from the ground state $\ket{G}$ to the defect state. The underlying idea is that the absorption from each QW will be proportional to the charge density $\rho_G(z_d)$. The spectral component of the absorption at the cantilever frequency $\omega_{m}$
will then be proportional to $\Xi(z_d)$ and thus to $\sigma^<$,  providing a measure of the CQED-induced ground-state excitation density. Note that the light-matter coupling in the QW is a collective phenomenon depending on the total plasma frequency of the QW $\omega_P$. This implies that while \Eq{ssmall} is verified we can safely neglect the change in the ground state energy due to transitions of $\sigma^<$ electrons to defect states \cite{Cwik2016}.

Using a simple Fermi golden rule approach we can write the total absorption rate as
\begin{align}
\label{abs}
\Gamma(t)&=\frac{4  P l_d n_{QW} \mu^2}{\hbar^2 \gamma_d c \sqrt{\epsilon_r} \epsilon_0}  \rho_G(z_d,t),
\end{align}
where $l_d\ll L_{QW}$ is the defect layer thickness, $\mu$ the defect dipole moment, and we 
explicitly marked the time-dependency of $\rho_G$ through the cavity length. 
Using \Eq{ssmall}, \Eq{sn}, and \Eq{L}, and assuming that the cantilever oscillations are small $a\ll1$, \Eq{rd} takes the form
\begin{align}
\label{rd}
\rho_G(z_d,t)&\approx\lvert \psi_1(z_d)\rvert^2 \sigma_e-
\Xi(z_d) \sigma^<(L_0)a \cos(\omega_{m} t).
\end{align}
The mirror oscillation thus translates in an oscillation of the charge coupled to the defect mode and, through \Eq{abs}, in an harmonic modulation of the laser absorption. 
Defining $\beta$ to be the quantum efficiency of the detector, the total photocurrent can be written as
\begin{align}
\label{IT}
I_{pc}(t)&=\beta e\left[\frac{ P}{\hbar\omega_d}-\Gamma(t)\right],
\end{align}
and the variation in the photocurrent due to the QWs 
\begin{align}
\label{Is}
I_s(t)&=\beta  e\Gamma(t).
\end{align}
This leads to the time-independent shot-noise current
\begin{align}
\label{Ish}
i_{sh}&= \sqrt{2eI_{pc}\Delta\nu},
\end{align}
where $\Delta\nu$ is the detection bandwidth.
Assuming the shot-noise to be the dominant noise term, that only a small fraction of the light is absorbed by the QW $I_s\ll I_{pc}$, and after having applied a pass-band filter around the frequency $\omega_m$ to remove the time-independent part of the signal, we can use \Eq{abs} to \Eq{Ish} to write the signal-to-noise ratio \cite{Wineland1987}
\begin{align}
\label{SNR}
SNR&=\frac{I_s}{i_{sh}}=\sqrt{\frac{2\hbar\omega_d P \beta}{\Delta\nu}} \frac{  l_d n_{QW} \mu^2  \Xi(z_d) a \sigma^<(L_0)}{\hbar^2 \gamma_d c \sqrt{\epsilon_r}\epsilon_0}.
\end{align}
It is worth to notice that the scheme presented above is solid against both fluctuations in the doping density $\sigma_e$, which gives only a continuum signal easily filtered out, and against spurious mirror-induced forces. This last point is quantitatively discussed in Appendix \ref{Casimir}, but it can be otherwise inferred by the fact that appreciable modifications of the electronic QW wavefunctions 
by metallic mirrors in close proximity would have otherwise became apparent in precise measurements of intersubband polariton systems with and without a metallic mirror \cite{anappara_signatures_2009}.

\section{Electrostatic measurement}
\label{Electrostatic}
\begin{figure}[t!]
\begin{center}
 \includegraphics[width=0.47\textwidth]{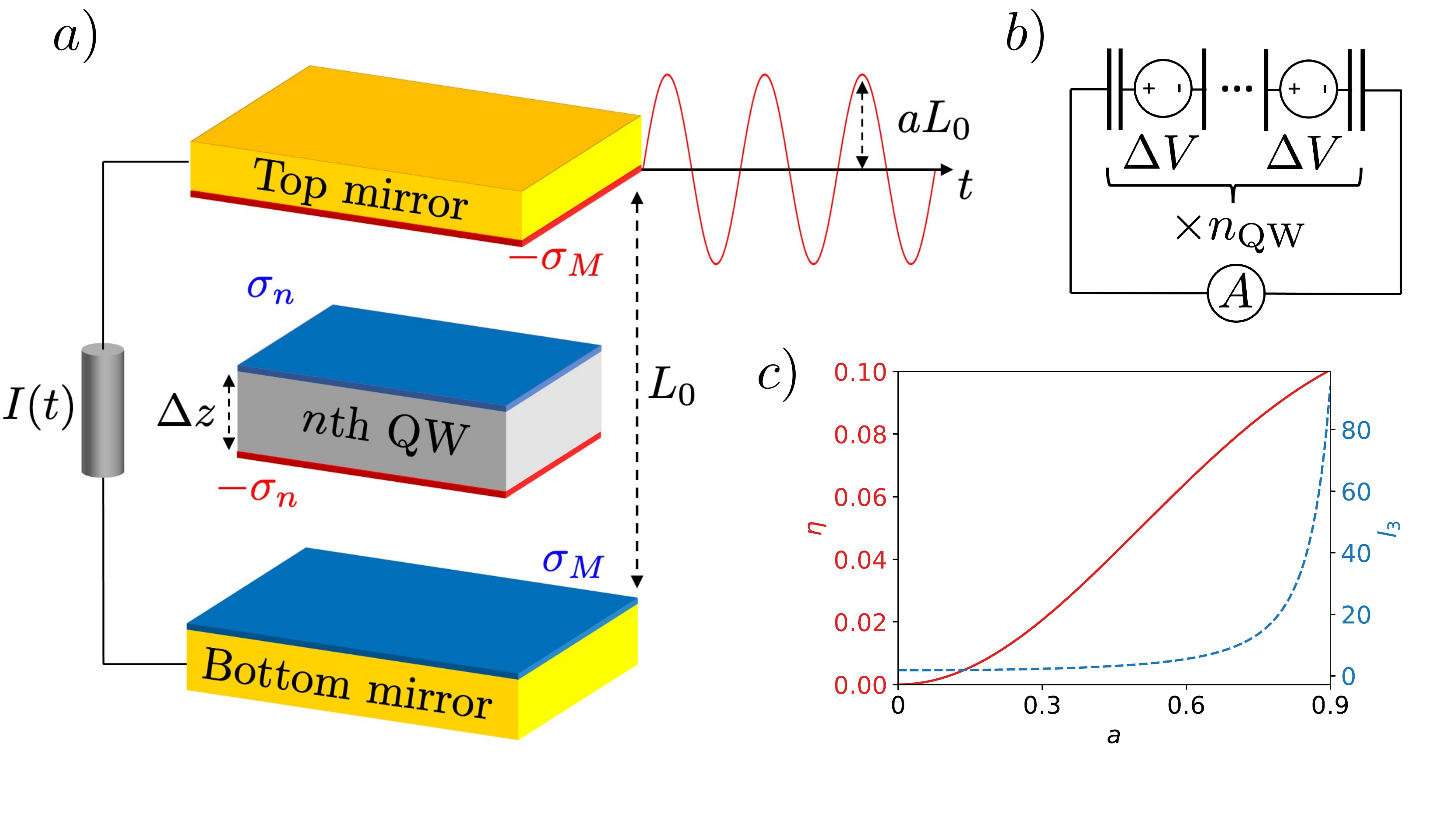}
\caption{Sketch of the setup described in Sec. \ref{Electrostatic} (a) and its schematic circuit representation (b). 
Each of the QWs, by creating a built-in potential due to the photon-induced charge displacement, generates a potential difference. When the position of the top mirror is modulated in time a current flows through the circuit. The definition of all the marked quantities can be found in the main text. 
In (c) we plot the discernability from \Eq{eta} (solid red) and them norm of the signal component from \Eq{norm} (dashed blue) as a function of the normalised oscillation amplitude $a$.
\label{Fig3}}
\end{center}
\end{figure}
In the previous Section we investigated the ground-state excited electronic population through a spectroscopic measurement of different charge distribution between the two subbands at a specific position $z_{d}$, quantified by the quantity $\Xi(z_d)$. Here instead we will discuss how to electrostatically measure the average charge displacement $\Delta z$ from \Eq{dz}. While the former procedure is arguably simpler to implement, being more robust against spurious effects, the latter one has the conceptual advantage to  measure a macroscopic real-space charge displacement, thus providing a more direct proof of the CQED-induced ground-state modification.

The scheme here proposed is based on the fact that resonator-induced charge perturbation described by the second term in \Eq{rz} creates a built-in electrostatic potential in each QW.  Considering a sample surface $S\gg L^2$ large enough to neglect border effects, such a potential can be determined solving the Poisson equation
\begin{align}
\label{Poisson}
\partial^2_z V(z)&=-\frac{e\sigma\Xi(z)}{\epsilon_0\epsilon_r}.
\end{align}
Writing a formal solution of \Eq{Poisson} 
\begin{align}
V(z)=-\int_{-\infty}^z dz' \int_{-\infty}^{z'} dz''\; \frac{e\sigma\Xi(z'')}{\epsilon_0\epsilon_r},
\end{align}
integrating by part the derivative of $V(z)$, and exploiting the charge conservation in each QW 
\begin{align}
\int_{\mathbb{R}}\Xi(z)dz=0,
\end{align}
we can calculate the potential drop across the QW
\begin{align}
\Delta V&=\frac{e\sigma}{\epsilon_0\epsilon_r}\int dz z \Xi(z).
\label{dV}
\end{align}
Using \Eq{rz} and \Eq{dz} we can put \Eq{dV} in the form
\begin{align}
\Delta V&=\frac{e\sigma\Delta z}{\epsilon_0\epsilon_r}.
\label{dV2}
\end{align}
From \Eq{dV2} we see that the voltage induced across each QW by the coupling with the transverse field of the resonator is the same created by two planes of electron density $\sigma$ at a distance $\Delta z$. Given that $\Delta z$ can be calculated or independently measured, a measure of $\Delta V$ will thus constitute a direct measure of the virtual excitation density in the ground state.

The induced potential difference can not be measured with a standard voltmeter because it is not a difference in the Fermi level but an in-built equilibrium potential. In can nevertheless be measured using a modified Kelvin probe \cite{melitz_kelvin_2011,Hutchison2013}. Kelvin probes are usually employed to measure the difference in work function between a known reference material and a sample, by using an AFM tip as the top plate of a capacitor whose lower plate is the sample. When the distance between the two is modulated by the tip oscillation, the capacitance also varies. Being the potential difference fixed by the work functions, a change in the capacitance leads to a change in the charge on the capacitor's plates, and thus to a measurable current. 
The apparatus we will consider, schematised in Fig. \ref{Fig3}(a), is thus similar to the one considered in the previous Section, with the top mirror now doubling as capacitor top plate. Moreover, the top and bottom mirrors are now electrically connected and the time-dependent current flowing between them is measured as the tip oscillates.

Applying the Kirchoff's law to the circuit equivalent of the apparatus shown in Fig. \ref{Fig3}(b) and exploiting \Eq{dV2} we can find the relation between the photon-induced electron density $\sigma$ and the electronic density on the mirrors' surfaces $\sigma_M$
\begin{align}
\label{znLM}
\Delta z n_{QW}\sigma + D &=L\sigma_M,
\end{align}
where we added a fixed dipole density $D$, modelling the in-built potentials independent of the mirror position which would be measured by a standard Kelvin probe.
Deriving over time the electron density on the mirror and using \Eq{ss} and \Eq{sGf3} we can then calculate the electric current  generated in the Kelvin probe by the tip oscillation
\begin{align}
\label{j}
I_K&=eS\dot{\sigma}_M=  eS\dot{L} \partial_L \frac{\Delta z n_{QW} \sigma+D}{L}\\
&=  eS\dot{L}\left[ \Delta z n_{QW} \partial_L\frac{\sigma^<}{L}- \frac{\tilde{D}}{L^2}\right].\nonumber
\end{align}
The first term in the square bracket in \Eq{j} represents the current induced by the virtual excitations caused by photonic modes confined by the cavity, and it is thus dependent on its length $L$, while in the second term we have grouped all the $L$-independent background polarizations 
by defining $\tilde{D}=\Delta z n_{QW} \sigma^>+D$.

Note that in \Eq{rd} only the term proportional to $\sigma^<$ is time-dependent, allowing for a simple suppression of the potentially much larger time-independent term proportional to $\sigma_e$. In \Eq{j} instead, both $\sigma^<$ and the background polarization $\tilde{D}$ lead to time-dependent signals. This is to be expected given that a standard Kelvin probe is meant to measure the static in-built potentials $\tilde{D}$, but it also implies extra care has to be paid in order to extract $\sigma^<$ from the measured signal. To this aim we can notice, plugging \Eq{sn} into \Eq{j} that the term due to CQED-induced charge modulation proportional to $\sigma^<$ scales as $L^{-3}$, while the background polarization as $L^{-2}$. Given that we know the time evolution of the cavity length $L(t)$ from \Eq{L}, it is {\it a priori} possible to distinguish the two contributions, but only if the tip explores a non-vanishing range of $L$, substantially changing the cavity length. This time we will thus not be able to limit ourselves to the perturbative regime $a\ll1$.

In order to filter out the background polarization we can introduce the two functions $g_p$ with $p=\{2,3\}$, periodic on the interval $\left[0,2\pi \right]$
\begin{align}
\label{ff}
g_p(x)&= \frac{\sin(x)}{\left[1+a\cos(x)\right]^p},
\end{align}
their $l^2$-norm 
\begin{align}
\label{norm}
l_p&=\sqrt{\int_0^{2\pi}  g_p(x)^2dx},
\end{align}
and their normalised versions $\tilde{g}_p(x)=g_p(x)/l_p$.
Substituting \Eq{sn} and \Eq{L} into \Eq{j} we can write the current as
\begin{align}
\label{jf}
I_K(t)&=I^< \tilde{g}_3(\omega_{m} t)+
\frac{eSa\omega_{m}\tilde{D}l_2}{L_0} \tilde{g}_2(\omega_{m} t),
\end{align}
with
\begin{align}
\label{Im}
I^<&=\frac{ Sa\omega_C\omega_{m} z_{12}^2\Delta z e^3 n_{QW}\sigma_e l_3}{2\pi \epsilon_0 c^2 \hbar L_0^2}. 
\end{align}
By projecting \Eq{jf} on the non-orthogonal normalised basis $\tilde{f}_p$ using the $l^2$ inner product we can then obtain a system of two equations in two unknowns which can be solved determining the coefficients, assuming their discernability 
\begin{align}
\label{eta}
\eta&=1-\lvert\int_0^{2\pi} dx \tilde{g}_2(x)\tilde{g}_3(x)\rvert^2,
\end{align}
is not vanishing.
In Fig.~\ref{Fig3}(c) we plot $\eta$ showing that for large oscillations ($a\approx 0.9$) we can reach a distinguishability of order $10\%$, while having a norm $l_3$ of order 100.
This procedure should allow us to measure the intensity of the current due to the cavity-induced ground-state modifications $I^<$ while filtering out static in-built potentials and all the other contributions not dependent on the position of the metallic mirror. The discrimination procedure could be further improved by performing control experiments either removing the bottom metallic mirror, or using doped semiconductors with lower plasma frequencies as oscillating mirrors, in order to obtain independent measurements of $\tilde{D}$.  Moreover, methods to analyse time-resolved data using non-orthogonal decompositions as the one we just discussed but integrating both known and unknown components also exist and could be exploited in this context although their analysis is beyond the scope of this paper \cite{Ki2019}.

Note that, beyond the doped QW system considered in this work, other asymmetric CQED platforms have been considered in the literature, both dielectric \cite{chestnov_terahertz_2017,DeLiberato2018} and superconducting \cite{garziano_vacuum-induced_2014}. It remains to be seen whether the electrostatic measurement scheme we introduced could be extended to these systems, further broadening our capability to measure and interact with the light-matter coupled ground state.

\section{Numerical results}
\label{Numerical}
In this Section we will estimate the magnitude of the SNR expected from \Eq{SNR} and of the current  from \Eq{Im}, with the objective to ascertain whether the two schemes we proposed are realisically implementable with present-day or near-future technology.

In order to determine a test QW structure, which for simplicity we will consider to be the same for the two schemes, we notice that strongly asymmetric potentials increase the average charge separation and thus $\Delta z$, but segregating the wavefunctions they also reduce the overlap between $\psi_1(z)$ and $\psi_2(z)$ and thus the dipole moment $z_{12}$. This same problem has been studied in the context of dipolar emission in driven asymmetric quantum wells, where the dipole $e\Delta z$ oscillates and causes  emission at the vacuum \cite{de_liberato_terahertz_2013} or pump \cite{kibis_matter_2009} Rabi-frequency. In Ref.~\cite{shammah_terahertz_2014} the parameter space of a simple asymmetric GaAs-based QW geometry has been explored in order to identify structures with both non-negligible $\Delta z$ and $z_{12}$. In the following we will use the parameters of a structure highlighted in such a publication, which are the ones we used in Fig.~\ref{Fig1}(a), to have comparatively large values of both figures of merit while being solid against fabrication tolerances. 

We thus consider $n_{QW}=10$ GaAs-based QWs  of length $L_{QW}=11.6$nm, with $\hbar\omega_{12}=125$meV, $z_{12}=0.18L_{QW}$, $\Delta z=0.11L_{QW}$, and the total QW stack height, including barriers, is taken to be $160$nm.
Considering a typical AFM scanning frequency $\omega_{m}=2\pi\times 100$kHz and an
equilibrium cavity length $L_0=2\mu$m, this makes it possible to choose values of the normalised oscillation amplitude $a$ up to $0.9$ before the tip touches the QW stack.
The QWs are each doped at a surface density $\sigma_{e}=10^{12}$cm$^{-2}$, corresponding to a Fermi energy $E_F\approx 35$meV from the bottom of the first subband, and we consider metallic mirrors with $\hbar\omega_C=10$eV. The corresponding cut-off wave vector $q_C=\frac{c}{\sqrt{\epsilon_r}\omega_C}$ is smaller than $0.5v_F$, assuring that \Eq{w12} remains at least qualitatively correct even close to the cut-off. 

We can thus numerically calculate the induced electronic excitation density per QW from \Eq{sn}, obtaining
$\sigma^<\approx 0.8\times 10^{6}$cm$^{-2} \ll \sigma_{e}$, justifying {\it a posteriori} our bosonic approach and our choice not to consider the backaction of the photon-induced charge displacement upon the electronic wavefunctions, which would have obliged us to solve a Schr\"odinger-Poisson equation instead of \Eq{Poisson}. 

For the sample to evaluate \Eq{SNR} we place the defect layer at $z_d=0.5L_{QW}$, corresponding to $\Xi(z_d)\approx \frac{2}{L_{QW}}$, and use a defect layer width of the order of the GaAs lattice parameter $l_d=0.5$nm, with a dipole moment $\mu=4$D, and a linewidth $\gamma_d=0.1$meV \cite{Khramatsov2016}. We consider a $100$ps plused laser with pulse energy of $1$mJ and repetition rate of $100$kHz, synced with the AFM tip having a small oscillation amplitude $a=0.1$. With a quantum efficiency $\beta=0.5$ and an integration time of ten minutes $\Delta\nu=\frac{\omega_m}{1200\pi}$ \Eq{SNR} leads to  $SNR\approx 1$. Given the realistic parameters used, and the fact the virtual electronic population in \Eq{sn} represents a lower estimate of the real value of $\sigma^<$, makes us confident that the spectroscopic measurement scheme we proposed is within technological reach.

In order to evaluate \Eq{Im}, we choose instead $a=0.7$ and consider a device surface $S=50\mu$m$^2$, leading to a lower bound of the expected current of the order of $I_K\approx 0.5$fA, small but still detectable with low-noise electronics. 
In this scheme the larger value of $a$ could {\it a priori} bring the mirror close enough to the QWs for the Casimir effect to exert a non-negligible force on the electron gas. While such a force remains very small, leading to a shift in the electronic position much smaller than $\Delta z$, it would affect all the electrons, potentially leading to an effect which could overshadow the larger shift of the few virtually-excited electrons.
While such contribution scales with higher powers of $L$, and it could thus potentially be filtered out extending the approach described in Sec.~\ref{Electrostatic}, in Appendix \ref{Casimir} we perform a rough overestimate of such an effect, showing that for the parameters chosen it should not pose a critical hurdle to the detection of CQED-induced effects.

Until now we completely neglected the effect of losses in our formalism. While the coupling with the environment will modify the population of virtual excitations in the ground state, in Ref.~\cite{de_liberato_virtual_2017} one of us demonstrated that such changes are limited even for over-damped systems.

\section{Conclusions}
In this paper we have explored the possibility of detecting resonator-induced effects on the electronic ground state of a CQED system. In particular we discussed two schemes which can achieve such an objective by providing a first direct measurement of virtual electronic excitations in the ground state of a CQED system. Instead of focussing, as most proposals do, on the detection of photonic excitations through non-adiabatic modulation, we aim at detecting material excitations using an adiabatic modulation. Our estimates show that detection of electronic CQED-induced modifications of the electronic ground state could be within technological reach. Although highly not trivial, the experiments we propose would not incur in the many hurdles linked with sub-cycle optical modulation and detection, which have until now thwarted multiple efforts to observe ground-state virtual photons through the non-adiabatic route.

We conclude by pointing out that the two schemes we proposed are not intended to represent an exhaustive list. The general theory developed in Sec. \ref{Theory} shows how the ground-state charge distribution in doped quantum wells is modified by the coupling with the electromagentic field. The results of many experiments which can be performed on QWs depend on different details of the ground-state charge distribution and we thus expect many different experiments could be used  to highlight its CQED-induced modifications. Other possible examples which are not investigated in this paper include nonlinear emission in asymmetric QWs \cite{Khurgin1989} or lifting of selection rules is symmetric ones.

We hope this work will stimulate novel interest toward the physics of virtual excitations and more broadly on the study of CQED effects on electronic degrees of freedom. 

\section*{Acknowledgements}
The authors acknowledge R. Colombelli and A. Tredicucci for useful discussions.
S.D.L. acknowledges support from a Royal Society Research Fellowship and from
the Philip Leverhulme Prize of the Leverhulme Trust. Y.W.'s studentship was financed by the Royal Society RGF\textbackslash EA\textbackslash 180062 grant.
\appendix

\section{Diagonalization of the PZW Hamiltonian}
Following Ref.~\cite{todorov_dipolar_2015}, the PZW Hamiltonian describing the system can be written in the bosonic approximation
\begin{align}
\label{H}
\hat{H}=\sum_{\mathbf{q}}\hbar \left(\hat{W}_{0\mathbf{q}}+\hat{W}_{I\mathbf{q}}
+\hat{W}_{D\mathbf{q}}\right),
\end{align}
with the first part describing the bare fields
\begin{align}
\label{W}
\hat{W}_{0\mathbf{q}}=& \sum_n \omega_{12} \hat{b}^{\dagger}_{n\mathbf{q}}\hat{b}_{n\mathbf{q}}
+ \sum_{m}\omega_{mq} \hat{a}^{\dagger}_{m\mathbf{q}}\hat{a}_{m\mathbf{q}},
\end{align}
the second their interaction
\begin{align}
\label{WI}
\hat{W}_{I\mathbf{q}}=&
 \sum_{mn}\omega_PK_{mnq}(\hat{a}^{\dagger}_{m\mathbf{-q}}+\hat{a}_{m\mathbf{q}})
(\hat{b}^{\dagger}_{n\mathbf{q}}+\hat{b}_{n\mathbf{-q}}),
\end{align}
and the third the  self-interaction of the electronic polarization
\begin{align}
\label{WD}
\hat{W}_{D\mathbf{q}}=&\sum_n\frac{\omega_{P}^2}{4\omega_{12}}
(\hat{b}^{\dagger}_{n\mathbf{q}}+\hat{b}_{n\mathbf{-q}})(\hat{b}_{n\mathbf{-q}}^{\dagger}+\hat{b}_{n\mathbf{q}}).
\end{align}
In \Eq{WI} we have the plasma frequency for a single QW 
\begin{eqnarray}
\label{OmegaP}
\omega_{P}^2&=&\frac{I_{12}\hbar e^2 \sigma_{e}}{2m_e^{*2} \epsilon_0\epsilon_r\omega_{12}},
\end{eqnarray}
with $m_e^*$ the effective mass of conduction electrons and the dimensionless coupling coefficient
\begin{align}
\label{Kmqn}
K_{mnq}^2=
\omega_{mq}\omega_{12}
\frac{z_{12}^2 m_e^{*2}}{ \hbar^2LI_{12}}\frac{2\cos^2 \theta_{mq}}{(1+\delta_{m0})} \cos^2(\frac{\pi m z_n}{L}),
\end{align}
parametrising the coupling between the bright matter mode in the $n$th QW  and the TM$_m$ photonic mode, where we defined $\theta_{mq}$ the angle of propagation relative to the $z$-axis
\begin{align}
\cos\theta_{mq}=\frac{cq}{\sqrt{\epsilon_r}\omega_{mq}}.
\end{align}

Note that the term in \Eq{WD}, quadratic in the plasma frequency $\omega_P$, is responsible for the phenomenon of polarization shift \cite{Geiser2012,de_liberato_quantum_2012}.
This can be seen by considering only the terms of the full Hamiltonian describing the bare field and the electron-electron interaction
\begin{align}
\label{We}
\hat{W}_{0\mathbf{q}}'=\hat{W}_{0\mathbf{q}}+\hat{W}_{D\mathbf{q}},
\end{align}
and performing a Bogoliubov rotation over the $\hat{b}_{n\mathbf{q}}$ operators, leading to a new Hamiltonian describing modified intersubband transitions operators $\hat{d}_{n\mathbf{q}}$ whose frequency is renormalised by the interaction
\begin{align}
\label{WR}
\hat{W}_{0\mathbf{q}}'=& \sum_n \sqrt{\omega_{12}^2+\omega_P^2}\, \hat{d}^{\dagger}_{n\mathbf{q}}\hat{d}_{n\mathbf{q}}
+ \sum_{m}\omega_{mq} \hat{a}^{\dagger}_{m\mathbf{q}}\hat{a}_{m\mathbf{q}}.
\end{align}

The electronic wavefunctions in the two subbands enter in play through the definition of the
 intersubband dipole
\begin{eqnarray}
\label{d}
z_{12}&=& \frac{\hbar}{2m_e^*\omega_{12}}\int  \left[\bar{\psi}_{1}(z)\partial_z \psi_{2}(z)- \psi_{2}(z)\partial_z\bar{\psi}_{1}(z)  \right]dz,\nonumber\\
\end{eqnarray}
and of the normalization factor
\begin{eqnarray}
\label{I12}
I_{12}&=&\int \left[ \bar{\psi}_{1}(z)\partial_z \psi_{2}(z)- \psi_{2}(z)\partial_z \bar{\psi}_{1}(z)  \right]^2dz.
\end{eqnarray}
In the bosonic regime we can then use the Hopfield-Bogoliubov approach to diagonalise $\hat{H}$ in \Eq{H} in terms of polaritonic modes 
\begin{eqnarray}
\label{pinaba}
\hat{p}_{s\mathbf{q}}&=\sum_m \left(x_{smq} \hat{a}_{m\mathbf{q}}+z_{smq} \hat{a}^{\dagger}_{m\mathbf{-q}}\right)\\&+\sum_n\left(
y_{snq}\hat{b}_{n\mathbf{q}}+w_{snq}\hat{b}_{n\mathbf{-q}}^{\dagger}\right),\nonumber
\end{eqnarray}
whose coupled ground state $\ket{G}$ is defined by $\hat{p}_{s\mathbf{q}}\ket{G}=0$.
The linear transformation in \Eq{pinaba} can then
be inverted to obtain
\begin{align}
\hat{b}_{n\mathbf{q}}&=\sum_s \bar{y}_{snq}\hat{p}_{s\mathbf{q}}-\bar{w}_{snq}\hat{p}^{\dagger}_{s\mathbf{q}}.
\end{align}

\section{Derivation of the bosonic expression for the electron density}
Here we will provide a derivation of the equation
\begin{align}
\label{rza}
\rho_G(z)-\rho_F(z)=&\frac{1}{S}\sum_{n\mathbf{q}}\left[\lvert\psi_2(z-z_n)\rvert^2-\lvert\psi_1(z-z_n)\rvert^2\right] \\&\times\nonumber
\bra{G}\hat{b}^{\dagger}_{n\mathbf{q}}\hat{b}_{n\mathbf{q}} \ket{G},
\end{align}
describing the change of the electronic density in the ground state created by the vacuum excitations. While physically intuitive its derivation requires some care as it links fermionic to bosonised quantities. We will thus expand the approach originally developed in Ref.~\cite{cortese_strong_2019} to calculate the electronic wavefunctions of photon-bound excitons.

We start by introducing the electron field operator projected on the first two subbands
\begin{align}
\hat{\Psi}(\mathbf{r})=\sum_{n\mathbf{k}}\left[\phi_{1n\mathbf{k}}(\mathbf{r})\hat{c}_{1n\mathbf{k}}+
\phi_{2n\mathbf{k}}(\mathbf{r})\hat{c}_{2n\mathbf{k}}\right].
\end{align}
Using as reference the electronic distribution in the absence of the resonator, we can then write the induced electron density as
\begin{align}
\rho_G(z)-\rho_F(z)&=\bra{G} \hat{\Psi}^{\dagger}(\mathbf{r})\hat{\Psi}(\mathbf{r}) \ket{G}-\bra{F} \hat{\Psi}^{\dagger}(\mathbf{r})\hat{\Psi}(\mathbf{r}) \ket{F}.
\label{dr}
\end{align}
The expression in \Eq{dr} can be simplified by noticing that both the free and coupled Hamiltonians, with ground states $\ket{F}$ and $\ket{G}$ respectively, commute with the parity of the total excitation number operator 
\begin{align}
\hat{T}&=\sum_{m\mathbf{q}}\hat{a}^{\dagger}_{m\mathbf{q}}\hat{a}_{m\mathbf{q}}+\frac{1}{2}\sum_{n\mathbf{k}}\left( \hat{c}^{\dagger}_{2n\mathbf{k}}\hat{c}_{2n\mathbf{k}}-\hat{c}^{\dagger}_{1n\mathbf{k}}\hat{c}_{1n\mathbf{k}}\right).
\end{align}
Both the coupled and free ground states have thus a well defined excitation number parity.
This implies that all the ground states expectation values 
involving terms which increase $\hat{T}$ by one, like  $\hat{c}^{\dagger}_{2n\mathbf{k}}\hat{c}_{1n'\mathbf{k}}$, have to vanish. Exploiting the fact that the number of electrons in each QW is fixed
\begin{align}
\sum_{\mathbf{k}}\left(\hat{c}^{\dagger}_{1n\mathbf{k}}\hat{c}_{1n\mathbf{k}}+\hat{c}^{\dagger}_{2n\mathbf{k}}\hat{c}_{2n\mathbf{k}}\right)&=\sigma_{e}S,
\end{align}
we can then put \Eq{dr} in the form
\begin{align}
\label{rzc}
\rho_G(z)-\rho_F(z)=&\frac{1}{S}\sum_{n\mathbf{k}}\left[\lvert\psi_2(z-z_n)\rvert^2-\lvert\psi_1(z-z_n)\rvert^2\right] \nonumber\\&\times
\bra{G}\hat{c}^{\dagger}_{2n\mathbf{k}}\hat{c}_{2n\mathbf{k}} \ket{G}.
\end{align}
In order to prove \Eq{rza} we have thus to demonstrate that
\begin{align}
\label{NFNB}
\bra{G}   \hat{N}_{Fn}\ket{G}
 &=
\bra{G}   \hat{N}_{Bn}\ket{G},
\end{align}
that is that the total number of electrons in the second subband of any QW
\begin{align}
\hat{N}_{Fn}=\sum_{\mathbf{k}}\hat{c}^{\dagger}_{2n\mathbf{k}}\hat{c}_{2n\mathbf{k}},
\end{align}
and the total number of matter excitations in the same QW
\begin{align}
\hat{N}_{Bn}=\sum_{\mathbf{q}}\hat{b}^{\dagger}_{n\mathbf{q}}\hat{b}_{n\mathbf{q}},
\end{align}
have the same expectation value in the coupled ground state. Note that by construction
\begin{align}
\label{NF}
\hat{N}_{Fn}\ket{F}&=\hat{N}_{Bn}\ket{F}=0,
\end{align}
and they thus also trivially coincide in the uncoupled ground state.
Using the definition of the intersubband transition operator in terms of electron operators
\begin{align}
\label{bnqa}
\hat{b}_{n\mathbf{q}}^{\dagger}&=\frac{1}{\sqrt{S\sigma_{e}}}\sum_{\mathbf{k}}\hat{c}^{\dagger}_{2n\mathbf{k+q}}\hat{c}_{1n\mathbf{k}},
\end{align}
we can verify that the fermionic commutator of $\hat{N}_{Fn}$ with the right-hand-side of \Eq{bnqa}
\begin{align}
\label{N2f}
\left[\hat{N}_{Fn},\hat{b}_{n\mathbf{q}}^{\dagger}\right]&=\hat{b}_{n\mathbf{q}}^{\dagger},
\end{align}
generates the same algebra as the commutator of $\hat{N}_{Bn}$ with the left-hand-side of the same equation, when considering the $\hat{b}_{n\mathbf{q}}$ as perfect bosons,
\begin{align}
\label{N2f}
\left[\hat{N}_{Bn},\hat{b}_{n\mathbf{q}}^{\dagger}\right]&=\hat{b}_{n\mathbf{q}}^{\dagger}.
\end{align}
Using the same bosonic assumption the coupled ground state $\ket{G}$ can be easily written in a perturbative expansion as a sum over orthogonal vectors $\ket{G}=\sum_{\zeta}\ket{\zeta}$, with
\begin{align}
\ket{\zeta}&=\chi_{\zeta}\prod_{j=1}^{j_{\zeta}} \hat{b}^{\dagger}_{\mathbf{q}_{\zeta j}}\prod_{h=1}^{h_{\zeta}} \hat{a}^{\dagger}_{m_{\zeta h}\mathbf{q}_{\zeta h}}\ket{F}.
\end{align}
Exploiting \Eq{NF} and the fact that $N_{Fn}$ and $N_{Bn}$ have the same commutation relations with all the operators appearing in the $\ket{\zeta}$ states, their expectation value in the ground state is necessarily the same, proving \Eq{NFNB} and thus concluding our proof of \Eq{rza}.\\

\section{Casimir-induced polarization}
\label{Casimir}
In the main body of the paper we considered two sources of induced polarization: the  virtual excitations due to the transverse electromagnetic mode confinement, leading to $\sigma_n\propto L^{-1}$, and in-built potentials independent from $L$.
At small scales there are nevertheless other interactions between the mirror and the electron gas which can lead to induced polarization and, should those forces be comparable or larger than the ones due to the transverse field, they would need to be taken into account when performing data analysis. 
In this Appendix we will derive the general form for the polarization induced on the $n$th QW from an arbitrary mirror-induced pressure $P(d_n)$ depending upon the distance between the mirror and the quantum well $d_n=L-z_n$.
We will also provide an upper estimate of such an effect for the Casimir force, showing that, although potentially measurable, such a force is not large enough to cover the electrostatic signal whose measurement we describe in Sec.~\ref{Electrostatic}.

A pressure $P(d_n)$ translates in a force per single electron $F(d_n)=P(d_n)/\sigma_e$ and its impact on one electron in the nth QW can thus be described by the Hamiltonian
\begin{align}
H_n&=F(d_n)\hat{z},
\end{align}
with $\hat{z}$ the electron coordinate.
To the lowest order the modified wavefunction in the first subband can thus be written as
\begin{align}
\ket{\psi_1'}&=\ket{\psi_1}-\frac{F(d_n)}{\hbar\omega_{12}}\bra{\psi_2}\hat{z}\ket{\psi_1}\ket{\psi_2},
\end{align}
and the corresponding shift in electron position is thus, to the lowest order
\begin{align}
\label{Dzp}
\Delta z'&=\bra{\psi_1'}\hat{z}\ket{\psi_1'}-\bra{\psi_1}\hat{z}\ket{\psi_1}\approx
\frac{2F(d_n) z_{12}^2}{\hbar\omega_{12}}.\end{align}
The total dipole density induced in the QW is thus 
\begin{align}
\label{Pnp}
P'_n=\sigma_e \Delta z',
\end{align}
which needs to be added to \Eq{znLM}.

While precise calculations of Casimir force in planar systems are possible \cite{Klimchitskaya2000} they are beyond the scope of the present paper. We can instead obtain an upper bound of such an effect by assuming that the electrons gas in a single QW is dense enough to act as a perfect mirror. Note that this would only be true for unphysical densities leading to coupling frequencies larger than $\omega_{12}$ as shown in Ref.~\cite{DeLiberato2014}. In the sample we are considering the effect of a single QW is instead largely perturbative and the result we will obtain only represents a very loose overestimate of the actual effect.
We can then consider each QW as a perfect mirror and use the analytical expression for the Casimir pressure between two planar metallic mirrors
\begin{align}
\label{Caspres}
P(d_n)&=-\frac{\hbar c \pi^2}{240\sqrt{\epsilon_r}d_n^4}.
\end{align}
Plugging \Eq{Caspres} into \Eq{Dzp} we obtain from \Eq{Pnp} the induced dipole, and using the parameters from Sec.~\ref{Numerical} we conclude that even at the point of the oscillation in which the mirror is the closest to the QW stack, and considering a QW on the very top of the stack, the induced dipole is only one-half of the CQED-induced one, dropping to $10^{-4}$ times smaller for a QW on the bottom of the stack when the mirror is at its furthest point. These estimates, as rough as they are, demonstrate that the Casimir effect, if not always completely negligible, does not represent a critical obstacle to the measurement scheme described in Sec.~\ref{Electrostatic} and {\it a fortiori} for the one from Sec.~\ref{Spectroscopic}.

\bibliographystyle{naturemag}
\bibliography{Bibliography}

\begin{thebibliography}{10}
\expandafter\ifx\csname url\endcsname\relax
  \def\url#1{\texttt{#1}}\fi
\expandafter\ifx\csname urlprefix\endcsname\relax\def\urlprefix{URL }\fi
\providecommand{\bibinfo}[2]{#2}
\providecommand{\eprint}[2][]{\url{#2}}

\bibitem{feist_polaritonic_2018}
\bibinfo{author}{Feist, J.}, \bibinfo{author}{Galego, J.} \&
  \bibinfo{author}{Garcia-Vidal, F.~J.}
\newblock \bibinfo{title}{Polaritonic {Chemistry} with {Organic} {Molecules}}.
\newblock \emph{\bibinfo{journal}{ACS Photonics}} \textbf{\bibinfo{volume}{5}},
  \bibinfo{pages}{205--216} (\bibinfo{year}{2018}).

\bibitem{ballarini_polaritonics_2019}
\bibinfo{author}{Ballarini, D.} \& \bibinfo{author}{De~Liberato, S.}
\newblock \bibinfo{title}{Polaritonics: from microcavities to sub-wavelength
  confinement}.
\newblock \emph{\bibinfo{journal}{Nanophotonics}}  (\bibinfo{year}{2019}).

\bibitem{de_bernardis_cavity_2018}
\bibinfo{author}{De~Bernardis, D.}, \bibinfo{author}{Jaako, T.} \&
  \bibinfo{author}{Rabl, P.}
\newblock \bibinfo{title}{Cavity quantum electrodynamics in the nonperturbative
  regime}.
\newblock \emph{\bibinfo{journal}{Phys. Rev. A}} \textbf{\bibinfo{volume}{97}},
  \bibinfo{pages}{043820} (\bibinfo{year}{2018}).

\bibitem{frisk_kockum_ultrastrong_2019}
\bibinfo{author}{Frisk~Kockum, A.}, \bibinfo{author}{Miranowicz, A.},
  \bibinfo{author}{De~Liberato, S.}, \bibinfo{author}{Savasta, S.} \&
  \bibinfo{author}{Nori, F.}
\newblock \bibinfo{title}{Ultrastrong coupling between light and matter}.
\newblock \emph{\bibinfo{journal}{Nature Reviews Physics}}
  \textbf{\bibinfo{volume}{1}}, \bibinfo{pages}{19--40} (\bibinfo{year}{2019}).

\bibitem{forn-diaz_ultrastrong_2019}
\bibinfo{author}{Forn-Díaz, P.}, \bibinfo{author}{Lamata, L.},
  \bibinfo{author}{Rico, E.}, \bibinfo{author}{Kono, J.} \&
  \bibinfo{author}{Solano, E.}
\newblock \bibinfo{title}{Ultrastrong coupling regimes of light-matter
  interaction}.
\newblock \emph{\bibinfo{journal}{Rev. Mod. Phys.}}
  \textbf{\bibinfo{volume}{91}}, \bibinfo{pages}{025005}
  (\bibinfo{year}{2019}).

\bibitem{anappara_signatures_2009}
\bibinfo{author}{Anappara, A.~A.} \emph{et~al.}
\newblock \bibinfo{title}{Signatures of the ultrastrong light-matter coupling
  regime}.
\newblock \emph{\bibinfo{journal}{Phys. Rev. B}} \textbf{\bibinfo{volume}{79}},
  \bibinfo{pages}{201303} (\bibinfo{year}{2009}).

\bibitem{khurgin_excitonic_2001}
\bibinfo{author}{Khurgin, J.~B.}
\newblock \bibinfo{title}{Excitonic radius in the cavity polariton in the
  regime of very strong coupling}.
\newblock \emph{\bibinfo{journal}{Solid State Commun.}}
  \textbf{\bibinfo{volume}{117}}, \bibinfo{pages}{307} (\bibinfo{year}{2001}).

\bibitem{yang_verification_2015}
\bibinfo{author}{Yang, M.-J.}, \bibinfo{author}{Kim, N.~Y.},
  \bibinfo{author}{Yamamoto, Y.} \& \bibinfo{author}{Na, N.}
\newblock \bibinfo{title}{Verification of very strong coupling in a
  semiconductor optical microcavity}.
\newblock \emph{\bibinfo{journal}{New Journal of Physics}}
  \textbf{\bibinfo{volume}{17}}, \bibinfo{pages}{023064}
  (\bibinfo{year}{2015}).

\bibitem{brodbeck_experimental_2017}
\bibinfo{author}{Brodbeck, S.} \emph{et~al.}
\newblock \bibinfo{title}{Experimental {Verification} of the {Very} {Strong}
  {Coupling} {Regime} in a {GaAs} {Quantum} {Well} {Microcavity}}.
\newblock \emph{\bibinfo{journal}{Phys. Rev. Lett.}}
  \textbf{\bibinfo{volume}{119}}, \bibinfo{pages}{027401}
  (\bibinfo{year}{2017}).

\bibitem{khurgin_pliable_2018}
\bibinfo{author}{Khurgin, J.~B.}
\newblock \bibinfo{title}{Pliable polaritons: {Wannier} exciton-plasmon
  coupling in metal-semiconductor structures}.
\newblock \emph{\bibinfo{journal}{Nanophotonics}} \textbf{\bibinfo{volume}{8}},
  \bibinfo{pages}{629--639} (\bibinfo{year}{2018}).

\bibitem{galego_cavity-induced_2015}
\bibinfo{author}{Galego, J.}, \bibinfo{author}{Garcia-Vidal, F.~J.} \&
  \bibinfo{author}{Feist, J.}
\newblock \bibinfo{title}{Cavity-{Induced} {Modifications} of {Molecular}
  {Structure} in the {Strong}-{Coupling} {Regime}}.
\newblock \emph{\bibinfo{journal}{Phys. Rev. X}} \textbf{\bibinfo{volume}{5}},
  \bibinfo{pages}{041022} (\bibinfo{year}{2015}).

\bibitem{cwik_excitonic_2016}
\bibinfo{author}{Ćwik, J.~A.}, \bibinfo{author}{Kirton, P.},
  \bibinfo{author}{De~Liberato, S.} \& \bibinfo{author}{Keeling, J.}
\newblock \bibinfo{title}{Excitonic spectral features in strongly coupled
  organic polaritons}.
\newblock \emph{\bibinfo{journal}{Phys. Rev. A}} \textbf{\bibinfo{volume}{93}},
  \bibinfo{pages}{033840} (\bibinfo{year}{2016}).

\bibitem{cortese_collective_2017}
\bibinfo{author}{Cortese, E.}, \bibinfo{author}{Lagoudakis, P.~G.} \&
  \bibinfo{author}{De~Liberato, S.}
\newblock \bibinfo{title}{Collective {Optomechanical} {Effects} in {Cavity}
  {Quantum} {Electrodynamics}}.
\newblock \emph{\bibinfo{journal}{Phys. Rev. Lett.}}
  \textbf{\bibinfo{volume}{119}}, \bibinfo{pages}{043604}
  (\bibinfo{year}{2017}).

\bibitem{schafer_ab_2018}
\bibinfo{author}{Schäfer, C.}, \bibinfo{author}{Ruggenthaler, M.} \&
  \bibinfo{author}{Rubio, A.}
\newblock \bibinfo{title}{Ab initio nonrelativistic quantum electrodynamics:
  {Bridging} quantum chemistry and quantum optics from weak to strong
  coupling}.
\newblock \emph{\bibinfo{journal}{Phys. Rev. A}} \textbf{\bibinfo{volume}{98}},
  \bibinfo{pages}{043801} (\bibinfo{year}{2018}).

\bibitem{levinsen_microscopic_2019}
\bibinfo{author}{Levinsen, J.}, \bibinfo{author}{Li, G.} \&
  \bibinfo{author}{Parish, M.~M.}
\newblock \bibinfo{title}{Microscopic description of exciton-polaritons in
  microcavities}.
\newblock \emph{\bibinfo{journal}{Phys. Rev. Research}}
  \textbf{\bibinfo{volume}{1}}, \bibinfo{pages}{033120} (\bibinfo{year}{2019}).

\bibitem{ebbesen_hybrid_2016}
\bibinfo{author}{Ebbesen, T.~W.}
\newblock \bibinfo{title}{Hybrid {Light}-{Matter} {States} in a {Molecular} and
  {Material} {Science} {Perspective}}.
\newblock \emph{\bibinfo{journal}{Acc. Chem. Res.}}
  \textbf{\bibinfo{volume}{49}}, \bibinfo{pages}{2403} (\bibinfo{year}{2016}).

\bibitem{herrera_cavity-controlled_2016}
\bibinfo{author}{Herrera, F.} \& \bibinfo{author}{Spano, F.~C.}
\newblock \bibinfo{title}{Cavity-{Controlled} {Chemistry} in {Molecular}
  {Ensembles}}.
\newblock \emph{\bibinfo{journal}{Phys. Rev. Lett.}}
  \textbf{\bibinfo{volume}{116}}, \bibinfo{pages}{238301}
  (\bibinfo{year}{2016}).

\bibitem{galego_many-molecule_2017}
\bibinfo{author}{Galego, J.}, \bibinfo{author}{Garcia-Vidal, F.~J.} \&
  \bibinfo{author}{Feist, J.}
\newblock \bibinfo{title}{Many-{Molecule} {Reaction} {Triggered} by a {Single}
  {Photon} in {Polaritonic} {Chemistry}}.
\newblock \emph{\bibinfo{journal}{Phys. Rev. Lett.}}
  \textbf{\bibinfo{volume}{119}}, \bibinfo{pages}{136001}
  (\bibinfo{year}{2017}).

\bibitem{ruggenthaler_quantum-electrodynamical_2018}
\bibinfo{author}{Ruggenthaler, M.}, \bibinfo{author}{Tancogne-Dejean, N.},
  \bibinfo{author}{Flick, J.}, \bibinfo{author}{Appel, H.} \&
  \bibinfo{author}{Rubio, A.}
\newblock \bibinfo{title}{From a quantum-electrodynamical light–matter
  description to novel spectroscopies}.
\newblock \emph{\bibinfo{journal}{Nat Rev Chem}} \textbf{\bibinfo{volume}{2}},
  \bibinfo{pages}{0118} (\bibinfo{year}{2018}).

\bibitem{ribeiro_polariton_2018}
\bibinfo{author}{Ribeiro, R.~F.}, \bibinfo{author}{MartÃ­nez-MartÃ­nez,
  L.~A.}, \bibinfo{author}{Du, M.}, \bibinfo{author}{Campos-Gonzalez-Angulo,
  J.} \& \bibinfo{author}{Yuen-Zhou, J.}
\newblock \bibinfo{title}{Polariton chemistry: controlling molecular dynamics
  with optical cavities}.
\newblock \emph{\bibinfo{journal}{Chem. Sci.}} \textbf{\bibinfo{volume}{9}},
  \bibinfo{pages}{6325--6339} (\bibinfo{year}{2018}).

\bibitem{thomas_tilting_2019}
\bibinfo{author}{Thomas, A.} \emph{et~al.}
\newblock \bibinfo{title}{Tilting a ground-state reactivity landscape by
  vibrational strong coupling}.
\newblock \emph{\bibinfo{journal}{Science}} \textbf{\bibinfo{volume}{363}},
  \bibinfo{pages}{615} (\bibinfo{year}{2019}).

\bibitem{cortese_excitons_2020}
\bibinfo{author}{Cortese, E.} \emph{et~al.}
\newblock \bibinfo{title}{Excitons bound by photon exchange}.
\newblock \emph{\bibinfo{journal}{Nature Physics}}
  \textbf{\bibinfo{volume}{17}}, \bibinfo{pages}{31--35}
  (\bibinfo{year}{2021}).

\bibitem{ciuti_quantum_2005}
\bibinfo{author}{Ciuti, C.}, \bibinfo{author}{Bastard, G.} \&
  \bibinfo{author}{Carusotto, I.}
\newblock \bibinfo{title}{Quantum vacuum properties of the intersubband cavity
  polariton field}.
\newblock \emph{\bibinfo{journal}{Phys. Rev. B}} \textbf{\bibinfo{volume}{72}},
  \bibinfo{pages}{115303} (\bibinfo{year}{2005}).

\bibitem{de_liberato_virtual_2017}
\bibinfo{author}{De~Liberato, S.}
\newblock \bibinfo{title}{Virtual photons in the ground state of a dissipative
  system}.
\newblock \emph{\bibinfo{journal}{Nat. Commun.}} \textbf{\bibinfo{volume}{8}},
  \bibinfo{pages}{1465} (\bibinfo{year}{2017}).

\bibitem{benea-chelmus_electric_2019}
\bibinfo{author}{Benea-Chelmus, I.-C.}, \bibinfo{author}{Settembrini, F.~F.},
  \bibinfo{author}{Scalari, G.} \& \bibinfo{author}{Faist, J.}
\newblock \bibinfo{title}{Electric field correlation measurements on the
  electromagnetic vacuum state}.
\newblock \emph{\bibinfo{journal}{Nature}} \textbf{\bibinfo{volume}{568}},
  \bibinfo{pages}{202--206} (\bibinfo{year}{2019}).

\bibitem{de_liberato_electro-optical_2019}
\bibinfo{author}{De~Liberato, S.}
\newblock \bibinfo{title}{Electro-optical sampling of quantum vacuum
  fluctuations in dispersive dielectrics}.
\newblock \emph{\bibinfo{journal}{Phys. Rev. A}}
  \textbf{\bibinfo{volume}{100}}, \bibinfo{pages}{031801}
  (\bibinfo{year}{2019}).

\bibitem{lolli_ancillary_2015}
\bibinfo{author}{Lolli, J.}, \bibinfo{author}{Baksic, A.},
  \bibinfo{author}{Nagy, D.}, \bibinfo{author}{Manucharyan, V.~E.} \&
  \bibinfo{author}{Ciuti, C.}
\newblock \bibinfo{title}{Ancillary {Qubit} {Spectroscopy} of {Vacua} in
  {Cavity} and {Circuit} {Quantum} {Electrodynamics}}.
\newblock \emph{\bibinfo{journal}{Phys. Rev. Lett.}}
  \textbf{\bibinfo{volume}{114}}, \bibinfo{pages}{183601}
  (\bibinfo{year}{2015}).

\bibitem{de_liberato_quantum_2007}
\bibinfo{author}{De~Liberato, S.}, \bibinfo{author}{Ciuti, C.} \&
  \bibinfo{author}{Carusotto, I.}
\newblock \bibinfo{title}{Quantum {Vacuum} {Radiation} {Spectra} from a
  {Semiconductor} {Microcavity} with a {Time}-{Modulated} {Vacuum} {Rabi}
  {Frequency}}.
\newblock \emph{\bibinfo{journal}{Phys. Rev. Lett.}}
  \textbf{\bibinfo{volume}{98}}, \bibinfo{pages}{103602}
  (\bibinfo{year}{2007}).

\bibitem{carusotto_back-reaction_2012}
\bibinfo{author}{Carusotto, I.}, \bibinfo{author}{De~Liberato, S.},
  \bibinfo{author}{Gerace, D.} \& \bibinfo{author}{Ciuti, C.}
\newblock \bibinfo{title}{Back-reaction effects of quantum vacuum in cavity
  quantum electrodynamics}.
\newblock \emph{\bibinfo{journal}{Phys. Rev. A}} \textbf{\bibinfo{volume}{85}},
  \bibinfo{pages}{023805} (\bibinfo{year}{2012}).

\bibitem{stassi_spontaneous_2013}
\bibinfo{author}{Stassi, R.}, \bibinfo{author}{Ridolfo, A.},
  \bibinfo{author}{Di~Stefano, O.}, \bibinfo{author}{Hartmann, M.~J.} \&
  \bibinfo{author}{Savasta, S.}
\newblock \bibinfo{title}{Spontaneous {Conversion} from {Virtual} to {Real}
  {Photons} in the {Ultrastrong}-{Coupling} {Regime}}.
\newblock \emph{\bibinfo{journal}{Phys. Rev. Lett.}}
  \textbf{\bibinfo{volume}{110}}, \bibinfo{pages}{243601}
  (\bibinfo{year}{2013}).

\bibitem{garziano_switching_2013}
\bibinfo{author}{Garziano, L.}, \bibinfo{author}{Ridolfo, A.},
  \bibinfo{author}{Stassi, R.}, \bibinfo{author}{Di~Stefano, O.} \&
  \bibinfo{author}{Savasta, S.}
\newblock \bibinfo{title}{Switching on and off of ultrastrong light-matter
  interaction: {Photon} statistics of quantum vacuum radiation}.
\newblock \emph{\bibinfo{journal}{Phys. Rev. A}} \textbf{\bibinfo{volume}{88}},
  \bibinfo{pages}{063829} (\bibinfo{year}{2013}).

\bibitem{huang_photon_2014}
\bibinfo{author}{Huang, J.-F.} \& \bibinfo{author}{Law, C.~K.}
\newblock \bibinfo{title}{Photon emission via vacuum-dressed intermediate
  states under ultrastrong coupling}.
\newblock \emph{\bibinfo{journal}{Phys. Rev. A}} \textbf{\bibinfo{volume}{89}},
  \bibinfo{pages}{033827} (\bibinfo{year}{2014}).

\bibitem{cirio_multielectron_2019}
\bibinfo{author}{Cirio, M.}, \bibinfo{author}{Shammah, N.},
  \bibinfo{author}{Lambert, N.}, \bibinfo{author}{De~Liberato, S.} \&
  \bibinfo{author}{Nori, F.}
\newblock \bibinfo{title}{Multielectron {Ground} {State}
  {Electroluminescence}}.
\newblock \emph{\bibinfo{journal}{Phys. Rev. Lett.}}
  \textbf{\bibinfo{volume}{122}}, \bibinfo{pages}{190403}
  (\bibinfo{year}{2019}).

\bibitem{falci_ultrastrong_2019}
\bibinfo{author}{Falci, G.}, \bibinfo{author}{Ridolfo, A.},
  \bibinfo{author}{Di~Stefano, P.~G.} \& \bibinfo{author}{Paladino, E.}
\newblock \bibinfo{title}{Ultrastrong coupling probed by {Coherent}
  {Population} {Transfer}}.
\newblock \emph{\bibinfo{journal}{Scientific Reports}}
  \textbf{\bibinfo{volume}{9}}, \bibinfo{pages}{9249} (\bibinfo{year}{2019}).

\bibitem{dodonov_current_2010}
\bibinfo{author}{Dodonov, V.~V.}
\newblock \bibinfo{title}{Current status of the dynamical {Casimir} effect}.
\newblock \emph{\bibinfo{journal}{Phys. Scr.}} \textbf{\bibinfo{volume}{82}},
  \bibinfo{pages}{038105} (\bibinfo{year}{2010}).

\bibitem{wilson_observation_2011}
\bibinfo{author}{Wilson, C.~M.} \emph{et~al.}
\newblock \bibinfo{title}{Observation of the dynamical {Casimir} effect in a
  superconducting circuit}.
\newblock \emph{\bibinfo{journal}{Nature}} \textbf{\bibinfo{volume}{479}},
  \bibinfo{pages}{376} (\bibinfo{year}{2011}).

\bibitem{nation_colloquium_2012}
\bibinfo{author}{Nation, P.~D.}, \bibinfo{author}{Johansson, J.~R.},
  \bibinfo{author}{Blencowe, M.~P.} \& \bibinfo{author}{Nori, F.}
\newblock \bibinfo{title}{Colloquium: {Stimulating} uncertainty: {Amplifying}
  the quantum vacuum with superconducting circuits}.
\newblock \emph{\bibinfo{journal}{Rev. Mod. Phys.}}
  \textbf{\bibinfo{volume}{84}}, \bibinfo{pages}{1} (\bibinfo{year}{2012}).

\bibitem{gunter_sub-cycle_2009}
\bibinfo{author}{Günter, G.} \emph{et~al.}
\newblock \bibinfo{title}{Sub-cycle switch-on of ultrastrong light–matter
  interaction}.
\newblock \emph{\bibinfo{journal}{Nature}} \textbf{\bibinfo{volume}{458}},
  \bibinfo{pages}{178} (\bibinfo{year}{2009}).

\bibitem{halbhuber_non-adiabatic_2020}
\bibinfo{author}{Halbhuber, M.} \emph{et~al.}
\newblock \bibinfo{title}{Non-adiabatic stripping of a cavity field from
  electrons in the deep-strong coupling regime}.
\newblock \emph{\bibinfo{journal}{Nature Photonics}}
  \textbf{\bibinfo{volume}{14}}, \bibinfo{pages}{675--679}
  (\bibinfo{year}{2020}).

\bibitem{todorov_dipolar_2015}
\bibinfo{author}{Todorov, Y.}
\newblock \bibinfo{title}{Dipolar quantum electrodynamics of the
  two-dimensional electron gas}.
\newblock \emph{\bibinfo{journal}{Phys. Rev. B}} \textbf{\bibinfo{volume}{91}},
  \bibinfo{pages}{125409} (\bibinfo{year}{2015}).

\bibitem{de_bernardis_breakdown_2018}
\bibinfo{author}{De~Bernardis, D.}, \bibinfo{author}{Pilar, P.},
  \bibinfo{author}{Jaako, T.}, \bibinfo{author}{De~Liberato, S.} \&
  \bibinfo{author}{Rabl, P.}
\newblock \bibinfo{title}{Breakdown of gauge invariance in ultrastrong-coupling
  cavity {QED}}.
\newblock \emph{\bibinfo{journal}{Phys. Rev. A}} \textbf{\bibinfo{volume}{98}},
  \bibinfo{pages}{053819} (\bibinfo{year}{2018}).

\bibitem{de_liberato_stimulated_2009}
\bibinfo{author}{De~Liberato, S.} \& \bibinfo{author}{Ciuti, C.}
\newblock \bibinfo{title}{Stimulated {Scattering} and {Lasing} of
  {Intersubband} {Cavity} {Polaritons}}.
\newblock \emph{\bibinfo{journal}{Phys. Rev. Lett.}}
  \textbf{\bibinfo{volume}{102}}, \bibinfo{pages}{136403}
  (\bibinfo{year}{2009}).

\bibitem{grieser_depolarization_2016}
\bibinfo{author}{Grießer, T.}, \bibinfo{author}{Vukics, A.} \&
  \bibinfo{author}{Domokos, P.}
\newblock \bibinfo{title}{Depolarization shift of the superradiant phase
  transition}.
\newblock \emph{\bibinfo{journal}{Phys. Rev. A}} \textbf{\bibinfo{volume}{94}},
  \bibinfo{pages}{033815} (\bibinfo{year}{2016}).

\bibitem{cortese_strong_2019}
\bibinfo{author}{Cortese, E.}, \bibinfo{author}{Carusotto, I.},
  \bibinfo{author}{Colombelli, R.} \& \bibinfo{author}{{De Liberato}, S.}
\newblock \bibinfo{title}{Strong coupling of ionizing transitions}.
\newblock \emph{\bibinfo{journal}{Optica}} \textbf{\bibinfo{volume}{6}},
  \bibinfo{pages}{354--361} (\bibinfo{year}{2019}).

\bibitem{Marzin1989}
\bibinfo{author}{Marzin, J.-Y.} \& \bibinfo{author}{G\'erard, J.-M.}
\newblock \bibinfo{title}{Experimental probing of quantum-well eigenstates}.
\newblock \emph{\bibinfo{journal}{Phys. Rev. Lett.}}
  \textbf{\bibinfo{volume}{62}}, \bibinfo{pages}{2172--2175}
  (\bibinfo{year}{1989}).

\bibitem{Cwik2016}
\bibinfo{author}{\ifmmode~\acute{C}\else \'{C}\fi{}wik, J.~A.},
  \bibinfo{author}{Kirton, P.}, \bibinfo{author}{De~Liberato, S.} \&
  \bibinfo{author}{Keeling, J.}
\newblock \bibinfo{title}{Excitonic spectral features in strongly coupled
  organic polaritons}.
\newblock \emph{\bibinfo{journal}{Phys. Rev. A}} \textbf{\bibinfo{volume}{93}},
  \bibinfo{pages}{033840} (\bibinfo{year}{2016}).

\bibitem{Wineland1987}
\bibinfo{author}{Wineland, D.~J.}, \bibinfo{author}{Itano, W.~M.} \&
  \bibinfo{author}{Bergquist, J.~C.}
\newblock \bibinfo{title}{Absorption spectroscopy at the limit: detection of a
  single atom}.
\newblock \emph{\bibinfo{journal}{Opt. Lett.}} \textbf{\bibinfo{volume}{12}},
  \bibinfo{pages}{389--391} (\bibinfo{year}{1987}).

\bibitem{melitz_kelvin_2011}
\bibinfo{author}{Melitz, W.}, \bibinfo{author}{Shen, J.},
  \bibinfo{author}{Kummel, A.~C.} \& \bibinfo{author}{Lee, S.}
\newblock \bibinfo{title}{Kelvin probe force microscopy and its application}.
\newblock \emph{\bibinfo{journal}{Surface Science Reports}}
  \textbf{\bibinfo{volume}{66}}, \bibinfo{pages}{1--27} (\bibinfo{year}{2011}).

\bibitem{Hutchison2013}
\bibinfo{author}{Hutchison, J.~A.} \emph{et~al.}
\newblock \bibinfo{title}{Tuning the work-function via strong coupling}.
\newblock \emph{\bibinfo{journal}{Advanced Materials}}
  \textbf{\bibinfo{volume}{25}}, \bibinfo{pages}{2481--2485}
  (\bibinfo{year}{2013}).

\bibitem{Ki2019}
\bibinfo{author}{Ki, H.}, \bibinfo{author}{Lee, Y.}, \bibinfo{author}{Choi,
  E.~H.}, \bibinfo{author}{Lee, S.} \& \bibinfo{author}{Ihee, H.}
\newblock \bibinfo{title}{Svd-aided non-orthogonal decomposition (sanod) method
  to exploit prior knowledge of spectral components in the analysis of
  time-resolved data}.
\newblock \emph{\bibinfo{journal}{Structural Dynamics}}
  \textbf{\bibinfo{volume}{6}}, \bibinfo{pages}{024303} (\bibinfo{year}{2019}).

\bibitem{chestnov_terahertz_2017}
\bibinfo{author}{Chestnov, I.~Y.}, \bibinfo{author}{Shahnazaryan, V.~A.},
  \bibinfo{author}{Alodjants, A.~P.} \& \bibinfo{author}{Shelykh, I.~A.}
\newblock \bibinfo{title}{Terahertz {Lasing} in {Ensemble} of {Asymmetric}
  {Quantum} {Dots}}.
\newblock \emph{\bibinfo{journal}{ACS Photonics}} \textbf{\bibinfo{volume}{4}},
  \bibinfo{pages}{2726--2737} (\bibinfo{year}{2017}).

\bibitem{DeLiberato2018}
\bibinfo{author}{De~Liberato, S.}
\newblock \bibinfo{title}{Lasing from dressed dots}.
\newblock \emph{\bibinfo{journal}{Nature Photonics}}
  \textbf{\bibinfo{volume}{12}}, \bibinfo{pages}{4--6} (\bibinfo{year}{2018}).

\bibitem{garziano_vacuum-induced_2014}
\bibinfo{author}{Garziano, L.}, \bibinfo{author}{Stassi, R.},
  \bibinfo{author}{Ridolfo, A.}, \bibinfo{author}{Di~Stefano, O.} \&
  \bibinfo{author}{Savasta, S.}
\newblock \bibinfo{title}{Vacuum-induced symmetry breaking in a superconducting
  quantum circuit}.
\newblock \emph{\bibinfo{journal}{Phys. Rev. A}} \textbf{\bibinfo{volume}{90}},
  \bibinfo{pages}{043817} (\bibinfo{year}{2014}).

\bibitem{de_liberato_terahertz_2013}
\bibinfo{author}{De~Liberato, S.}, \bibinfo{author}{Ciuti, C.} \&
  \bibinfo{author}{Phillips, C.~C.}
\newblock \bibinfo{title}{Terahertz lasing from intersubband
  polariton-polariton scattering in asymmetric quantum wells}.
\newblock \emph{\bibinfo{journal}{Phys. Rev. B}} \textbf{\bibinfo{volume}{87}},
  \bibinfo{pages}{241304} (\bibinfo{year}{2013}).

\bibitem{kibis_matter_2009}
\bibinfo{author}{Kibis, O.~V.}, \bibinfo{author}{Slepyan, G.~Y.},
  \bibinfo{author}{Maksimenko, S.~A.} \& \bibinfo{author}{Hoffmann, A.}
\newblock \bibinfo{title}{Matter {Coupling} to {Strong} {Electromagnetic}
  {Fields} in {Two}-{Level} {Quantum} {Systems} with {Broken} {Inversion}
  {Symmetry}}.
\newblock \emph{\bibinfo{journal}{Phys. Rev. Lett.}}
  \textbf{\bibinfo{volume}{102}}, \bibinfo{pages}{023601}
  (\bibinfo{year}{2009}).

\bibitem{shammah_terahertz_2014}
\bibinfo{author}{Shammah, N.}, \bibinfo{author}{Phillips, C.~C.} \&
  \bibinfo{author}{De~Liberato, S.}
\newblock \bibinfo{title}{Terahertz emission from ac {Stark}-split asymmetric
  intersubband transitions}.
\newblock \emph{\bibinfo{journal}{Phys. Rev. B}} \textbf{\bibinfo{volume}{89}},
  \bibinfo{pages}{235309} (\bibinfo{year}{2014}).

\bibitem{Khramatsov2016}
\bibinfo{author}{Khramtsov, E.~S.} \emph{et~al.}
\newblock \bibinfo{title}{Radiative decay rate of excitons in square quantum
  wells: Microscopic modeling and experiment}.
\newblock \emph{\bibinfo{journal}{Journal of Applied Physics}}
  \textbf{\bibinfo{volume}{119}}, \bibinfo{pages}{184301}
  (\bibinfo{year}{2016}).

\bibitem{Khurgin1989}
\bibinfo{author}{Khurgin, J.}
\newblock \bibinfo{title}{Second-order intersubband nonlinear-optical
  susceptibilities of asymmetric quantum-well structures}.
\newblock \emph{\bibinfo{journal}{J. Opt. Soc. Am. B}}
  \textbf{\bibinfo{volume}{6}}, \bibinfo{pages}{1673--1682}
  (\bibinfo{year}{1989}).

\bibitem{Geiser2012}
\bibinfo{author}{Geiser, M.} \emph{et~al.}
\newblock \bibinfo{title}{Ultrastrong coupling regime and plasmon polaritons in
  parabolic semiconductor quantum wells}.
\newblock \emph{\bibinfo{journal}{Phys. Rev. Lett.}}
  \textbf{\bibinfo{volume}{108}}, \bibinfo{pages}{106402}
  (\bibinfo{year}{2012}).

\bibitem{de_liberato_quantum_2012}
\bibinfo{author}{De~Liberato, S.} \& \bibinfo{author}{Ciuti, C.}
\newblock \bibinfo{title}{Quantum theory of intersubband polarons}.
\newblock \emph{\bibinfo{journal}{Phys. Rev. B}} \textbf{\bibinfo{volume}{85}},
  \bibinfo{pages}{125302} (\bibinfo{year}{2012}).

\bibitem{Klimchitskaya2000}
\bibinfo{author}{Klimchitskaya, G.~L.}, \bibinfo{author}{Mohideen, U.} \&
  \bibinfo{author}{Mostepanenko, V.~M.}
\newblock \bibinfo{title}{Casimir and van der waals forces between two plates
  or a sphere (lens) above a plate made of real metals}.
\newblock \emph{\bibinfo{journal}{Phys. Rev. A}} \textbf{\bibinfo{volume}{61}},
  \bibinfo{pages}{062107} (\bibinfo{year}{2000}).

\bibitem{DeLiberato2014}
\bibinfo{author}{De~Liberato, S.}
\newblock \bibinfo{title}{Light-matter decoupling in the deep strong coupling
  regime: The breakdown of the purcell effect}.
\newblock \emph{\bibinfo{journal}{Phys. Rev. Lett.}}
  \textbf{\bibinfo{volume}{112}}, \bibinfo{pages}{016401}
  (\bibinfo{year}{2014}).

\end{thebibliography}

\end{document}